\documentclass{emulateapj}

\newcommand{\gr}{$\gamma$-ray \,}
\newcommand{\grs}{$\gamma$-rays \,}
\shorttitle{Gamma-ray emission of SN~1006}
\shortauthors{Berezhko et al.}

\begin{document}

\title{Nonthermal emission of supernova remnant SN~1006 revisited: theoretical
  model and the H.E.S.S. results}

\author{E.G. Berezhko\altaffilmark{1},
        L.T. Ksenofontov\altaffilmark{1},
    and H.J. V\"olk\altaffilmark{2}
}

\altaffiltext{1}{Yu.G. Shafer Institute of Cosmophysical Research and Aeronomy,
                     31 Lenin Ave., 677980 Yakutsk, Russia}
\altaffiltext{2}{Max Planck Institut f\"ur Kernphysik,
                Postfach 103980, D-69029 Heidelberg, Germany}

\email{berezhko@ikfia.ysn.ru}

\begin{abstract}
  The properties of the Galactic supernova remnant (SNR) SN~1006 are
  theoretically re-analyzed in the light of the recent
  H.E.S.S. results. Nonlinear kinetic theory is used to determine the momentum
  spectrum of cosmic rays (CRs) in space and time in the supernova remnant
  SN~1006. The physical parameters of the model -- proton injection rate,
  electron-to-proton ratio and downstream magnetic field strength -- are
  determined through a fit of the result to the observed spatially-integrated
  synchrotron emission properties. The only remaining unknown astronomical
  parameter, the circumstellar gas number density, is determined by a
  normalization of the amplitude of the gamma-ray flux to the observed
  amplitude. The bipolar morphology of both nonthermal X-ray and \gr emissions
  is explained by the preferential injection of suprathermal nuclei and
  subsequent magnetic field amplification in the quasi-parallel regions of the
  outer supernova shock. The above parameters provide an
    improved fit to all existing nonthermal emission data, including the TeV
  emission spectrum recently detected by H.E.S.S., with the circumstellar
  hydrogen gas number density $N_\mathrm{H}\approx 0.06~\mbox{cm}^{-3}$ close
  to values derived from observations of thermal X-rays. The hadronic and
  leptonic \gr emissions are of comparable strength. The overall energy of
  accelerated CRs at the present epoch is of the order of 5 \% of the total
  hydrodynamic explosion energy, and is predicted to rise with time by a factor
  of $\approx 2$. The relevance of CR escape from the SNR for the spectrum of
  the \gr emission is demonstrated. The sum of the results suggests that
  SN~1006 is a CR source with a high efficiency of nuclear CR production, as
  required for the Galactic CR sources, both in flux as well as in cutoff
  energy.
\end{abstract}


\section{Introduction}

The Galactic cosmic rays (CRs), with proton energies below a few $~10^{15}$~eV,
are generally believed to be accelerated in shell-type supernova remnants
(SNRs). To establish that the Galactic SNRs are
indeed the main sources of the Galactic CRs one needs at least a handful of
SNRs with clearly determined astronomical parameters, like the type of
supernova explosion, the SNR age, the distance, and the properties of the
circumstellar medium, together with the nonthermal emission characteristics
from which the energetic particle spectra can be deduced with a theoretical
model. The most suitable objects are type Ia supernovae which have rather
old and low-mass progenitors, not associated with star forming regions, and
thus supposedly explode into a uniform circumstellar medium -- at least into
one which is not modified by mass loss in the form of a stellar wind. The
type of explosion also restricts the total explosion energy rather tightly
and fixes the ejecta mass to a value close to the Chandrasekhar mass.

One such object is the historical remnant SN~1006: the distance was determined
using optical measurements with relatively high precision \citep{winkler03} and
all other astronomical parameters are quite well known
\citep[e.g.][]{cassam08}. A special characteristic of SN~1006 is the fact
that it lies about 550 pc above the Galactic plane \citep{bocchino11} in a
very low-density environment. This unusual property suggests that the ambient
interstellar medium (ISM) is also free of density inhomgeneities from cooling
instabilities \citep[e.g.][]{inoue12}.

It has been shown earlier \citep{bkv09}, in the following referred to as BKV09,
that a nonlinear kinetic theory of CR acceleration in SNRs \citep{bek96,bv97}
\footnote{ Similar approaches were recently developed and applied to study the
  properties of other SNRs \citep{za10,zp11}.}  was consistent
with all observational data available at the time. This included the
H.E.S.S. energy flux density $\Phi(\epsilon_1, \epsilon_2) = 2.5 \times
10^{-12}$~erg cm$^{-2}$ s$^{-1}$ of the Very High Energy (VHE:$>100$~GeV) \gr
emission from one of the detected polar caps, integrated over the observed
energy interval $\epsilon_1 < \epsilon < \epsilon_2$, where $\epsilon_1 =
0.2$~TeV and $\epsilon_2 = 40$~TeV \citep{naumann08}, as well as the fluxes of
the nonthermal emission in the radio and X-ray regions measured with Chandra
\citep{allen04} and Suzaku \citep{bamba08}. Note that the above H.E.S.S. energy
flux density $\Phi(\epsilon_1, \epsilon_2)$ was attributed in BKV09 to the
whole remnant, whereas in fact it represents about half of the total flux. This
is the reason why the values of the ambient ISM density and of the SN explosion
energy, estimated below, are correspondingly larger in comparison with BKV09.

Compared with BKV09, in the following the physical parameters of SN~1006 --
mainly the ambient gas density and the energy production efficiency -- are
re-estimated, based on the more recent measurements of the \gr spectrum and of
the emission morphology \citep{sn1006hess10}. In addition, the time
profile of the acceleration efficieny is presented and the radial dependence
of the \gr brightness components is given. Finally the radial profiles of the
gas density and of the CR spectral energy density per unit logarithmic
bandwidth are compared in a discussion of the energetic particle escape from
the remnant.

\section{ Summary of  assumptions and earlier results }

As a type Ia supernova SN~1006 presumably ejects roughly a Chandrasekhar mass
$M_\mathrm{ej}=1.4
M_{\odot}$. Since the gas density is observed to vary only
mildly across the SNR \citep{acero07}, it appears reasonable to assume the
circumstellar gas density and magnetic field to be roughly
uniform. The ISM mass density $\rho_0=m_\mathrm{p}N_\mathrm{ISM} \approx
1.4m_\mathrm{p}N_\mathrm{H}$, is characterized by the hydrogen number density
$N_\mathrm{H}$. As the most reliable distance estimate the value $d=2.2$~kpc is
adopted \citep{winkler03}.

As in BKV09 a nonlinear kinetic theory of CR acceleration in
SNRs \citep{bek96,bv97} is applied. With one
exception this theory includes equations for the most important physical
processes which influence CR acceleration and SNR dynamics: shock modification
by CR backreaction, MHD wave damping and thus gas heating within the shock
transition, and synchrotron losses of CR electrons under the assumption of
  a strongly amplified magnetic field within the remnant $B(t)$.

The values of three scalar parameters in the
governing equations (proton injection rate $\eta$, electron to proton ratio
$K_\mathrm{ep}$ and the upstream magnetic field strength 
$B_0$) can be determined from a fit of the solutions that contain these
parameters to the observed spatially integrated synchrotron emission data
at the present epoch. Both, $\eta$ and $K_\mathrm{ep}$, are assumed to be
independent of time. The parameter values for SN~1006,
evaluated in this way, agree very well with the Chandra measurements of the
X-ray synchrotron filaments and were obtained in BKV09 by the analysis of the
radio data compiled by \citet{allen04} \citep[see also][]{allen01,allen08} and
of the most accurate X-ray data of Chandra \citep{allen04} and Suzaku
\citep{bamba08}.

\section{Results }

In order to explain the detailed \gr spectrum, values of the
hydrodynamic supernova explosion energy $E_\mathrm{sn}= 2.4\times 10^{51}$~erg
and $E_\mathrm{sn}= 1.9\times 10^{51}$~erg are taken to fit the observed shock
size $R_\mathrm{s}=9.5\pm0.35$~pc and shock speed $V_\mathrm{s}=4500\pm
1300$~km s$^{-1}$ \citep{moffett93,katsuda09} at the current epoch
$t_{SN}\approx 10^3$~yr for the ISM hydrogen number densities
$N_\mathrm{H}=0.08~\mbox{cm}^{-3}$ and $N_\mathrm{H}=0.05~\mbox{cm}^{-3}$,
respectively. These densities are consistent with the observed level of the VHE
emission, as shown below. The best-fit value of the upstream magnetic field
strength $B_0=30$~$\mu$G is quite insensitive to $N_\mathrm{H}$. The
resulting current total shock compression ratios $\sigma$ for
$N_\mathrm{H}=0.08~\mbox{cm}^{-3}$ and $N_\mathrm{H}=0.05~\mbox{cm}^{-3}$
become now $\sigma = 5.1$~and 4.9, respectively, whereas the
subshock compression ratios $\sigma_\mathrm{s}$ remain both
close to $\sigma_\mathrm{s}=3.7$.
Note, that the explosion energy value $E_\mathrm{sn}= 2.4\times 10^{51}$~erg
is somewhat lower compared with $E_\mathrm{sn}= 3\times 10^{51}$~erg that one
would expect extrapolating the value $E_\mathrm{sn}= 3.8\times 10^{51}$~erg
obtained earlier \citep{kbv05} for $N_\mathrm{H}=0.1~\mbox{cm}^{-3}$ according
to the expected dependence $E_\mathrm{sn}\propto N_\mathrm{H}$.
However this difference is well within the range determined by
the uncertenties of the observed values $R_\mathrm{s}$ and $V_\mathrm{s}$
taking into account the relation $R_\mathrm{s}\propto E_\mathrm{sn}^{1/5}$.

Since the properties of  the accelerated CR nuclear and electron spectra
and their dependence on the relevant physical parameters, as well as the
dynamical properties of the system were described in detail in BKV09, they will
not be discussed here once more.

\subsection{Acceleration efficiency}
\begin{figure}
 \epsscale{0.8}
\plotone{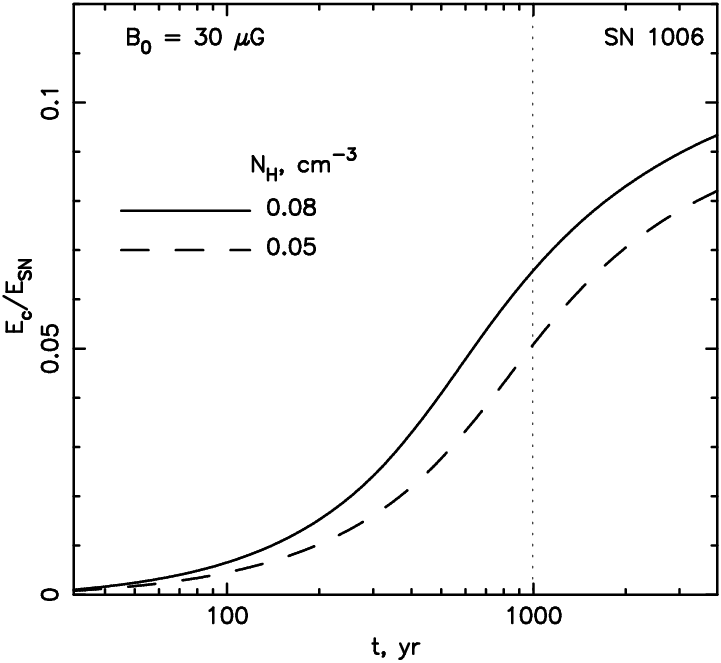}
\caption{Overall energy $E_\mathrm{c}$ of accelerated CRs, normalized to
    the total hydrodynamic energy release $E_\mathrm{sn}$, as a function of
  time, calculated for $N_\mathrm{H}=0.08~\mbox{cm}^{-3}$ ({\it solid line})
  and for $N_\mathrm{H}=0.05~\mbox{cm}^{-3}$ ({\it dashed line}).  
Here and
    in all following figures the quantity $B_0 = B_\mathrm{d}/\sigma$
    denotes the amplified upstream magnetic field strength, where
    $\sigma$ is the overall shock compression ratio and $B_\mathrm{d}$
is the downstream magnetic field strength. The {\it vertical dotted
      line} shows the current evolutionary stage $t_\mathrm{SN}$.}
\label{fig1}
\end{figure}

In Fig.~\ref{fig1} the time dependence of the fractional energy
$E_\mathrm{c}/E_\mathrm{sn}$ contained in accelerated CRs during the SNR
evolution is presented.  Note that the value of $E_\mathrm{c}$ is reduced by a
factor of $f_\mathrm{re} = 0.2$ (see below) compared with the value calculated
within the spherically symmetric model.  According to Fig.~\ref{fig1},
$E_\mathrm{c}/E_{sn} \approx 0.05$ and 0.065 for
$N_\mathrm{H}=0.05~\mbox{cm}^{-3}$ and $N_\mathrm{H}=0.08~\mbox{cm}^{-3}$,
respectively. This is lower than the value $\approx 0.1$ which is required on
average for each SNR for their population to be the main source of CRs in the
Galaxy.  The reason is that SN~1006 is a quite young object in an evolutionary
sense: according to Fig.~\ref{fig1} $E_\mathrm{c}(t)/E_\mathrm{sn} $ is
expected to approach this canonical value in the subsequent evolution. On
the other hand, the requirement for an average efficiency of $\approx 0.1$ is
based on an assumed Galactic average value $E_\mathrm{sn} = 10^{51}$~erg
which is about one half of our value $E_\mathrm{sn} =
2x10^{51}$~erg. Therefore the calculated efficiency $E_\mathrm{c}/E_{sn}
\approx 0.05$ fulfills the average requirement even for the present epoch.
%

\subsection{Overall nonthermal spectra}
\begin{figure*}
\plotone{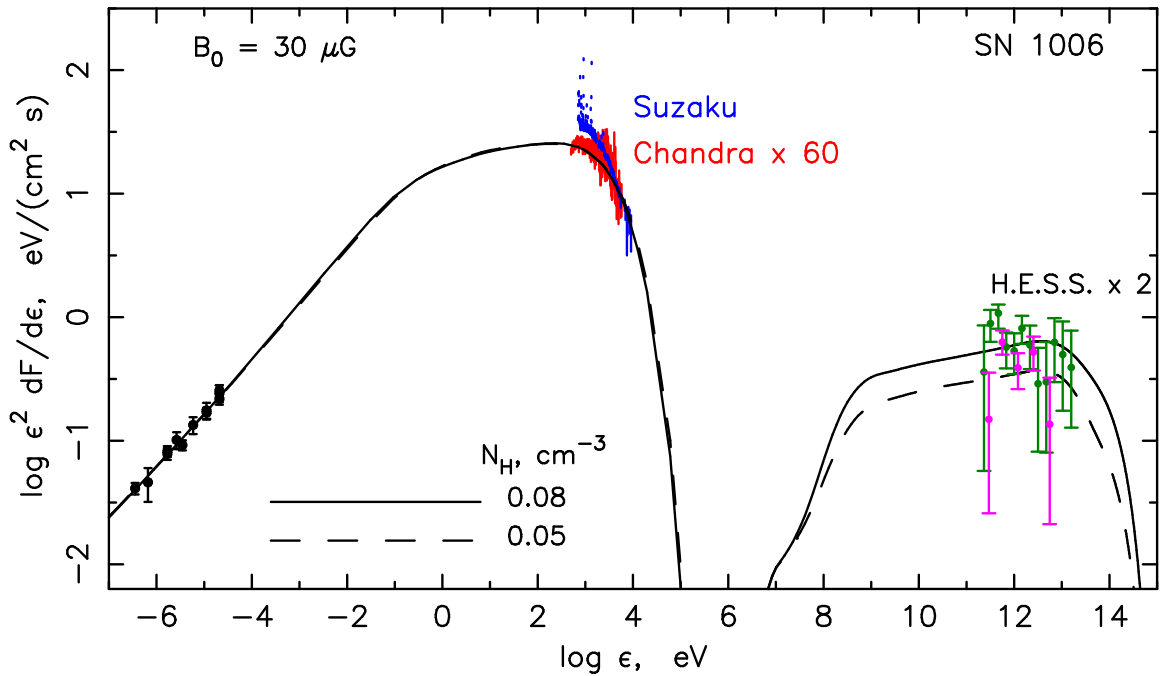}
\caption{Spatially integrated spectral energy distribution of SN~1006,
  calculated for $N_\mathrm{H}=0.08~\mbox{cm}^{-3}$ ({\it solid line}) and
  $N_\mathrm{H}=0.05~\mbox{cm}^{-3}$ ({\it dashed line}), and compared with
  observational data.  The observed radio spectrum is from a compilation of
  \citet{allen04}. The form of the Chandra X-ray spectrum was measured for a
  small region of the bright northeastern (NE) rim of SN~1006 \citep{allen04},
  whereas the Suzaku spectrum ({\it blue color}) \citep{bamba08} is for the
  entire remnant. The corresponding Chandra X-ray flux has been multiplied 
    by a factor of 60 ({\it red colour}) such as to be consistent with the
  Suzaku flux for energies $\epsilon > 2$~keV. The H.E.S.S. data
  \cite{sn1006hess10} are for the NE region ({\it green colour}) and the SW
  region ({\it magenta colour}).  Both these H.E.S.S. data sets have been
    multiplied by a factor of 2, in order to facilitate a comparison with
    the global
    theoretical spectra.}
\label{fig2}
\end{figure*}

Fig.~\ref{fig2} illustrates the consistency of the synchrotron and \gr spectra,
calculated with the best set of parameters ( $\eta = 2\times 10^{-4}$ as well
as $K_\mathrm{ep} = 4.5\times 10^{-4}$ for $N_\mathrm{H}=0.05~\mbox{cm}^{-3}$,
and $\eta = 2\times 10^{-4}$ as well as $K_\mathrm{ep} = 3.2\times 10^{-4}$ for
$N_\mathrm{H}=0.08~\mbox{cm}^{-3}$), with the observed spatially integrated
spectra.  The H.E.S.S. data \citep{sn1006hess10} for the NE and SW limbs,
respectively, have been multiplied by a factor of 2, in order facilitate
comparison with the full deduced \gr flux.  As can be seen from
Fig.~\ref{fig2}, the calculated synchrotron spectrum fits the observations both
in the radio and the X-ray ranges \citep{allen04,bamba08} very well.  Note that
the Chandra flux \citep{allen04} is from a small region of the bright NE limb
with minimal contributions from thermal X-rays; this X-ray flux was normalized
to the overall Suzaku flux \citep{bamba08} at energies $\epsilon > 2$~keV (for
details, see BKV09).

The only important parameter which can not be determined from the analysis of
the synchrotron emission data is the external gas number density
$N_\mathrm{H}$: Fig.~\ref{fig2} shows that the spectrum of synchrotron emission
is almost non-sensitive to the ambient gas density.  Consequently, numerical
solutions have been calculated for the pair of values
$N_\mathrm{H}=0.08~\mbox{cm}^{-3}$ and $N_\mathrm{H}=0.05~\mbox{cm}^{-3}$ which
appear to bracket the density range consistent with the H.E.S.S. \gr
measurements. It should be noted that this estimate for $N_\mathrm{H}$ is by a
factor of $\approx 1.5$ larger than the estimate of BKV09. The reason is that
in the analysis of BKV09 the total calculated \gr energy flux was compared with
the energy flux from the NE rim. However, the latter flux corresponded only to
about 50\% of the total observed energy flux, as it had been reported for the
H.E.S.S. instrument by \citet{naumann08}. A detailed comparison of the \gr
spectrum with the calculated spectrum will be made in the context of
Fig.~\ref{fig4}.
%

\subsection{Gamma-ray morphology}
The \gr morphology, as found in the H.E.S.S. measurements \citep{sn1006hess10},
is consistent with the prediction of a polar cap geometry on account of the
strongly preferred injection of nuclear particles in the quasi-parallel regions
of the shock \citep{vbk03}. Such a geometry has also been found experimentally
from an analysis of the synchrotron morphology in hard X-rays by \citet{ro04}
and \citet{cassam08}. This means that the \gr emission calculated in the
spherically symmetric model must be renormalized (reduced) by a factor
$f_\mathrm{re} \approx 0.2$, as in \citet{kbv05} and BKV09.  This
renormalization factor is applied here.  The total \gr flux from the source is
the sum of the fluxes from these two regions.

This morphology is also a key argument for the existence of an energetically
dominant nuclear CR component in SN~1006, because only such a component can
amplify the magnetic field to the observed degree. The accelerated electrons
are unable to amplify the magnetic field to the required level (BKV09).

The question, whether these bright NE and SW regions of SN~1006 represent
quasi-parallel portions of the SN shock or not, is still debated
\citep{petruk09,schneiter10,morlino10}. On the other hand \citet{bocchino11}
have recently argued from the radio morphology that the radio limbs --
morphologically similar to the X-ray and \gr limbs -- are polar caps and that
electrons are accelerated with quasi-parallel injection efficieny. Like the
aforementioned theoretical arguments and X-ray analyses, this argues against a
scenario, where the magnetic field is perpendicular to the shock normal
\citep{fulbright90}. \citet{bocchino11} also concluded that the remaining
asymetries (converging limbs and different surface brightness), clearly visible
also in the H.E.S.S. \gr data \citep{sn1006hess10}, could be explained by a
gradient of the ambient ISM magnetic field strength.

Additional arguments which  confirm that the NE and SW regions
correspond to quasi-parallel shock regions are given by
\citet{reynoso11}. Their analysis of the polarization of radio emission led
them to the conclusion that the magnetic field in SN~1006 is radial at the NE
and SW lobes but tangential in the SE of the radio shell. In addition they
established the maximum fractional polarization in the SE which implies that
the magnetic field is highly ordered there. On the other hand the low
fractional polarization in the lobes suggests a considerable randomization of
the magnetic field.  The fact that a turbulent magnetic field coexists with the
brightest synchrotron emission in the SW strongly supports efficient
acceleration of the CR nuclear component, followed by magnetic field
amplification which provides its own directional randomization.

\subsection{Relative contributions to the gamma-ray flux}
\begin{figure}[t]
 \plotone{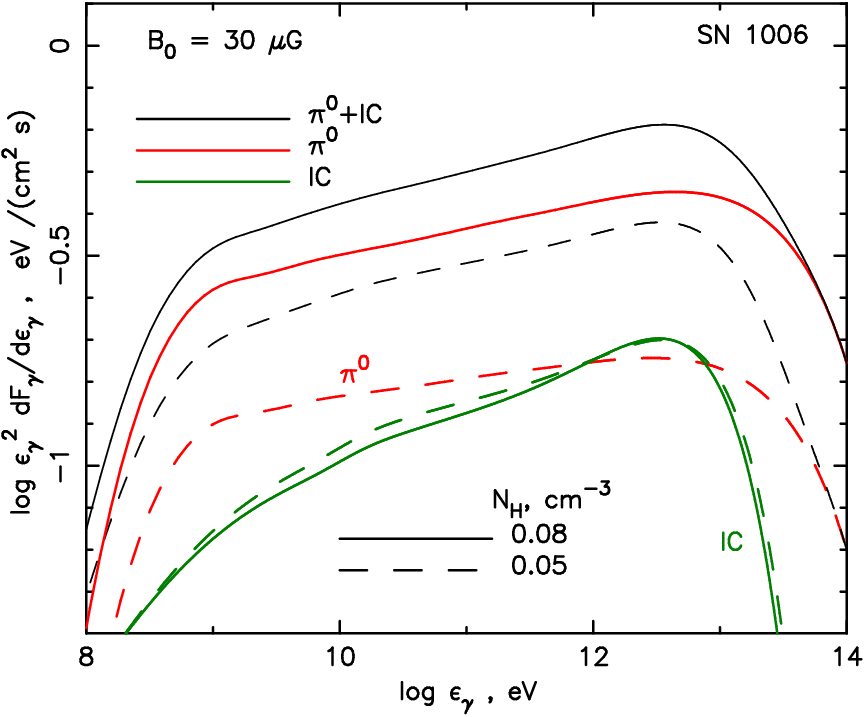}
 \caption{Total ($\pi^0$-decay + IC) ({\it black lines}), $\pi^0$-decay ({\it
     red lines}) and IC ({\it green lines}) differential $\gamma$-ray energy
   fluxes as a function of $\gamma$-ray energy, calculated for the ISM hydrogen
   number densities $N_\mathrm{H}=0.05~\mbox{cm}^{-3}$ ({\it dashed lines}) and
   $N_\mathrm{H}=0.08~\mbox{cm}^{-3}$ ({\it solid lines}), respectively, and
   for the injection parameters derived from the fit of the overall
   synchrotron spectrum.}
\label{fig3}
\end{figure}

Fig.~\ref{fig3} shows the $\pi^0$-decay, the inverse Compton (IC), and the
total ($\pi^0$-decay plus IC) $\gamma$-ray energy spectra of the remnant,
calculated for $N_\mathrm{H}=0.05~\mbox{cm}^{-3}$ and
$N_\mathrm{H}=0.08~\mbox{cm}^{-3}$, respectively. It can be seen that the \gr
spectrum produced by the nuclear CRs is rather close to the IC emission
spectrum produced by CR electrons alone, especially in the case of
$N_\mathrm{H}=0.05~\mbox{cm}^{-3}$. Even for the case
  $N_\mathrm{H}=0.035~\mbox{cm}^{-3}$ which is close to the minimum density
  $N_\mathrm{H}=0.03~\mbox{cm}^{-3}$ for the remnant, as argued by
  \citet{acero07},\, the total predicted \gr flux is still quite similar in
  spectral shape to the total flux for $N_\mathrm{H}=0.05~\mbox{cm}^{-3}$ (and
  essentially also for $N_\mathrm{H}=0.08~\mbox{cm}^{-3}$), as shown in Fig.6
  of BKV09. Unless the actual gas density would be small compared to
  $N_\mathrm{H}=0.03~\mbox{cm}^{-3}$, it would therefore be very difficult to
  observationally distinguish a hypothetical dominant IC scenario from the
  mixed \gr spectrum discussed here.  Therefore, in the VHE range -- except
perhaps in the last decade of energy, near the cutoff, see subsection 3.6
-- the observed \gr spectrum alone is not able to discriminate between the
hadronic $\pi^0$-decay and the leptonic IC \gr components. However, it was
already shown by \citet{kbv05} that such a low total VHE emission flux, with a
highly depressed IC \gr flux due to the synchrotron losses of high energy
electrons, is only possible if the nuclear CR component is efficiently produced
with accompanying strong magnetic field amplification.

\subsection{Radial profile of gamma-ray brightness}
A possibility for an experimental discrimination between $\pi^0$-decay and IC
\grs is given by the measurement of the {\it radial profile} of the \gr
brightness.  It was shown early on \citep{bkv02} that the radial profile of the
TeV $\pi^0$-decay \gr emission has a peak near the remnant rim with a width of
about 20\% of the remnant radius. On the other hand, the peak of the IC \gr
emission was predicted to be much narrower, actually narrower by an order of
magnitude.  This latter characteristic is physically identical to the
filamentary structure detected subsequently with Chandra in keV X-rays
\citep{bamba03,long03} (and for Cas~A by \citet{Vink03}), because the keV
synchrotron emission and the TeV \gr IC emission are produced by the same
electrons, with energies approaching 100~TeV. 

\begin{figure}[t]
 \plotone{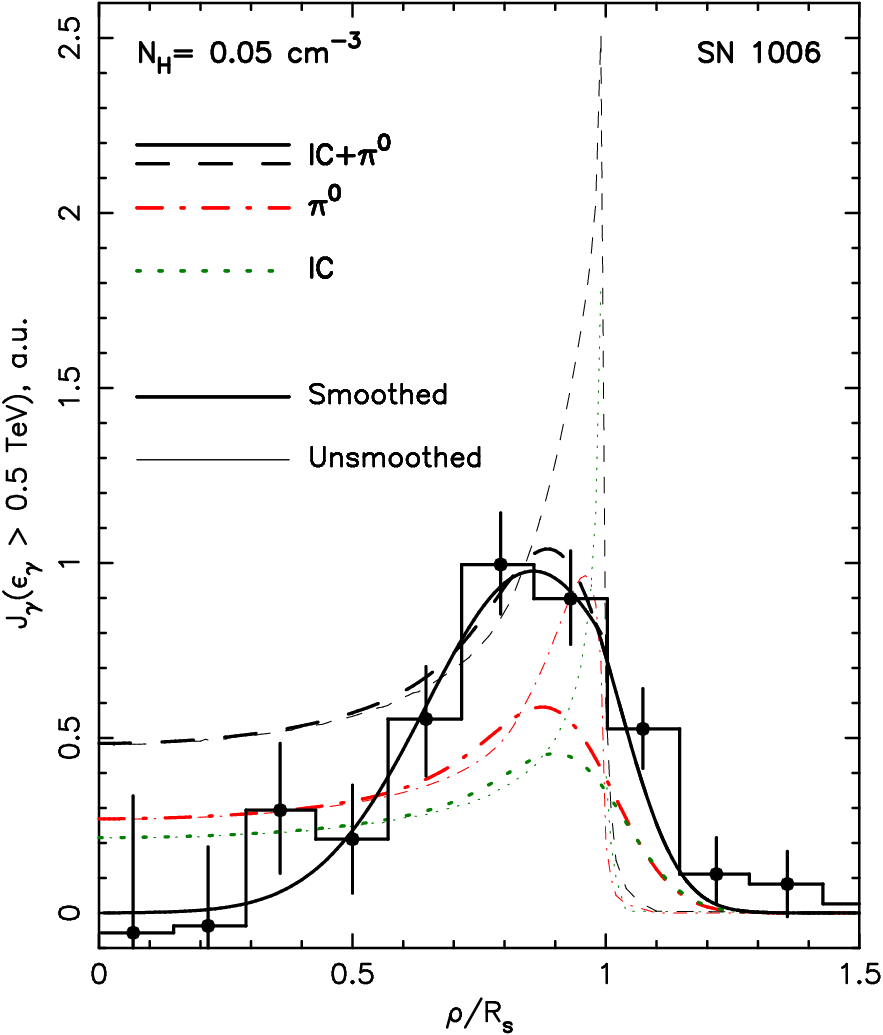}
 \caption{Radial dependence of the\gr brightness for \gr energies
   $\epsilon_{\gamma}>0.5$~TeV. IC ({\it green color}) and $\pi^0$-decay ({\it
     red color}) \gr components together with their sum are shown by {\it thin
     dotted, thin dash-dotted} and {\it thin dashed} lines respectively. These
   radial profiles, smoothed to the H.E.S.S. point spread function, are shown
   by corresponding {\it thick} lines. The {\it thick solid} line corresponds
   to the total \gr emission that originates in two cone regions.  The
   H.E.S.S. measurements are shown as well \citep{sn1006hess10}.  The 
     angular radius 0.24$^{\circ}$ corresponding to the position of the peak of
   the H.E.S.S. \gr profile is taken equal to
   $\rho=0.85R_s$.}
\label{fig4}
\end{figure}

In Fig.~\ref{fig4} radial profiles of the \gr brightness $J_{\gamma}(\rho)$~are
shown as functions of radial projection distance $\rho$ corresponding to
energies $\epsilon_{\gamma}>0.5$~TeV, calculated for
$N_\mathrm{H}=0.05~\mbox{cm}^{-3}$.  It is seen that the profile of the IC
component is indeed dramatically thinner than the profile of the $\pi^0$-decay
component. The profile of the total \gr emission, which is the
sum of these two components, has a width at half maximum of about $0.15R_s$
that is smaller than that observed by H.E.S.S.  The total radial emission
profile, smoothed to the H.E.S.S. point spread function, agrees with the
H.E.S.S. observations for $\rho > 0.5 R_s$ but exceeds the H.E.S.S. values
considerably at low radial distances $\rho < 0.5 R_s$.  The reason for this
discrepancy is simple. As it has been said in subsection 3.3, only those
nuclear CRs that occupy two cones with opening angle 20$^{\circ}$, with
symmetry axes going from the remnant center toward the NE and SW directions,
respectively, should be taken into account to model the hadronic emission of
the actual remnant due to the strong angular dependence of the nuclear
injection rate.  The radial profiles of the \gr emission lobes, corresponding
to the NE-SW directions produced by these CRs, are well consistent with the
H.E.S.S. data (Fig.~\ref{fig4}).  In this case the line of sight at
$\rho<0.7R_s$ does not intersect the outer spherical region of the remnant
which contains most of the accelerated CRs that provide the brightness
depression at $\rho < 0.7 R_s$.

A discrimination between these
$\pi^0$-decay and IC \gr components would require a \gr
instrument with an angular resolution that is an order of magnitude higher than
that of H.E.S.S., or of any other existing \gr instrument. It is,
however to be noted that the measured radial H.E.S.S. profile, which gives a
width of the measured peaks of 20\% of the remnant radius \citep{sn1006hess10},
is clear evidence that the nuclear CR component is accelerated
efficiently. Indeed, if the energy content in CR nuclei were small due to
inefficient proton acceleration one would have to conclude that the magnetic
field in the remnant is not amplified to any substantial degree.  In such an
inefficient scenario electrons with energies of tens of TeV, which produce the
TeV \gr IC emission, would be extremely smoothly distributed across the remnant
-- almost uniformly in the spherically symmetric case
\citep{bkv02,bkv03}. Therefore the expected radial width of the \gr emission
region would be considerably larger than that expected from efficiently
accelerated protons.

\subsection{External density}
\begin{figure}[t]
 \plotone{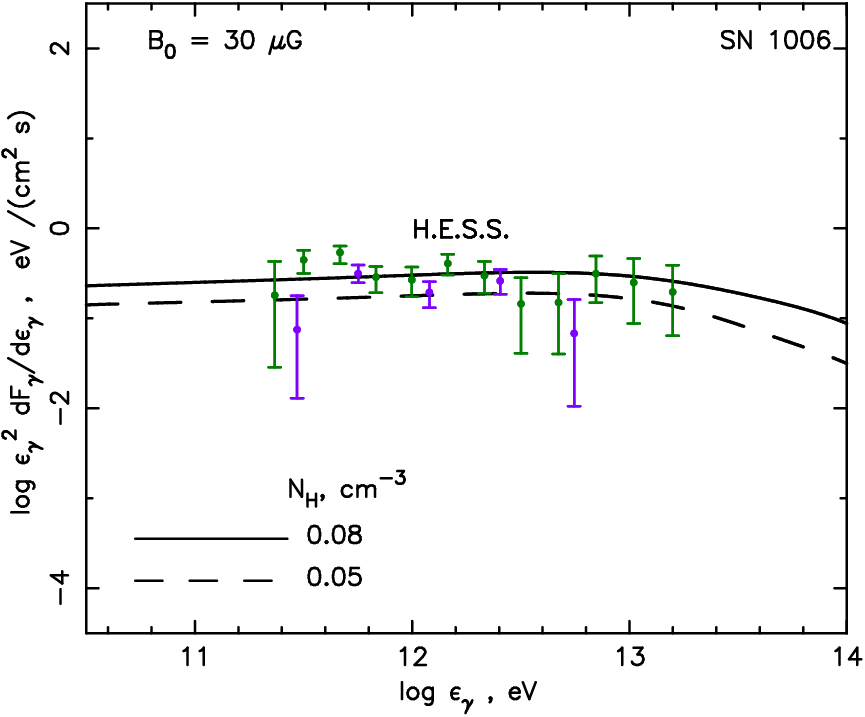}
 \caption{Spatially integrated \gr energy flux from SN 1006, divided by a
     factor of 2, calculated for the two ISM hydrogen number densities
   $N_\mathrm{H}=0.05~\mbox{cm}^{-3}$ ({\it dashed lines}) and
   $N_\mathrm{H}=0.08~\mbox{cm}^{-3}$ ({\it solid lines}), respectively. The
   H.E.S.S. data \citep{sn1006hess10} for the NE region ({\it green colour}) and
the
   SW region ({\it magenta colour}) are shown as well.}
\label{fig5}
\end{figure}

In order to exhibit the most detailed comparison of the calculated \gr spectra
with the observations, in Fig.~\ref{fig5} the theoretical spectra for two
ambient densities are presented together with the observed H.E.S.S. spectra for
each of the polar caps on an expanded energy scale. For ease of comparison, the
theoretical spectra are divided by a factor of two, which would correspond to
the emission from a single polar cap region for an ideal dipolar morphology.

The H.E.S.S. spectra from the two regions are compatible with decreasing
power-law distributions  $dF_{\gamma}/d\epsilon_{\gamma}\propto
  \epsilon_{\gamma}^{-\Gamma}$, with $\Gamma \simeq 2.3$ and somewhat different
  fluxes $\Phi ( >1\mathrm{TeV})$, in the detected energy range $11.4 \leq
\log \epsilon_{\gamma} \leq 13.2$ \citep{sn1006hess10}. As 
  Fig.~\ref{fig2}, Fig.~\ref{fig3} and Fig.~\ref{fig5} show, 
  the H.E.S.S. data are also compatible with the more complex total 
  theoretical \gr spectral energy  flux density over this range that has
a maximum at $\log \epsilon_{\gamma} \approx 12.6$. Above 10 GeV the
theoretical spectral energy flux can be approximated analytically 
by a power law with an exponential cutoff:
\[ 
dF_{\gamma}/d\epsilon_{\gamma}\propto 
\epsilon_{\gamma}^{-1.9} \exp{(-\epsilon_{\gamma}/\epsilon_\mathrm{c})}, 
\] with cutoff energies $\epsilon_\mathrm{c}=30$~TeV and 37~TeV for
$N_\mathrm{H}=0.05~\mbox{cm}^{-3}$ and $N_\mathrm{H}=0.08~\mbox{cm}^{-3}$,
respectively.  These two analytical curves represent the best fit to the
  H.E.S.S. data. At the same time they
  coincide within 4\%, respectively, with the 
  theoretical curves presented in Fig.~\ref{fig5}.

The quality of fit to  the H.E.S.S. data is characterized by the values
$\chi^2/dof=1.05$ and 2.2 for  the NE and SW part of  the remnant
respectively. It is comparable with the quality of fit by  the pure power
law  distributions 
 $dF_{\gamma}/d \epsilon_{\gamma}\propto \epsilon_{\gamma}^{-\Gamma}$, 
with $\Gamma \simeq
  2.3$ and different fluxes  \citep{sn1006hess10}, which have, according to
our calculation,  a $\chi^2/dof=1.3$ and 2.5  for the two regions,
respectively.

In order to discriminate between those different spectral forms the
observational energy range should be extended both below the present
H.E.S.S. threshold energy and above the present maximum energy detected by
H.E.S.S.

According to Fig.~\ref{fig5} the H.E.S.S. data are consistent with an ISM
number density in the interval $0.05\le N_\mathrm{H}\le 0.08~\mbox{cm}^{-3}$.
There is also some indication that the actual number density is closer to the
lower end of this interval, say $N_\mathrm{H}\approx 0.06~\mbox{cm}^{-3}$.
However this is a weak conclusion, given the approximations made in the
theoretical model in terms of the reduction factor $f_\mathrm{re}$ (see
above). It should be noted nevertheless that the lower end of this interval is
also preferred from the point of view of supernova explosion theory: the
corresponding hydrodynamic explosion energy $E_\mathrm{sn}\approx 1.9\times
10^{51}$~erg is close to the upper end of the typical range of type Ia SN
explosion energies that vary by a factor of about two
\citep{gamezo05,brs06}. In addition, from their X-ray measurements also
\citet{acero07} concluded that a value $N_\mathrm{H}\approx
0.05~\mbox{cm}^{-3}$ is probably representative for the ambient density around
SN~1006.

A solid observational conclusion is that the total emission from the NE limb
exceeds that from the SW limb \citep{sn1006hess10}. Given that the hadronic \gr
flux is proportional to $N_\mathrm{H}^2$, the result in Fig.5 suggests that
this excess is due to a gas density difference $\delta N_\mathrm{H} \sim 0.02$
cm$^{-3}$ between the two \gr emission regions across the SNR, i.e. a relative
difference of $\sim 30\%$. Given that the density difference between the
thermally emitting NW and SE regions is of the order of $0.1$ cm$^{-3}$
\citep{acero07} such a minor density difference is indeed quite
plausible. However, it does not explain the deviations from full dipolar
symmetry in terms of a noticeable convergence of the limbs and thus should
simply complement the gradient in the magnetic field strength claimed by
\citet{bocchino11} from radio observations.

\subsection{Escape}
\begin{figure}[t]
 \plotone{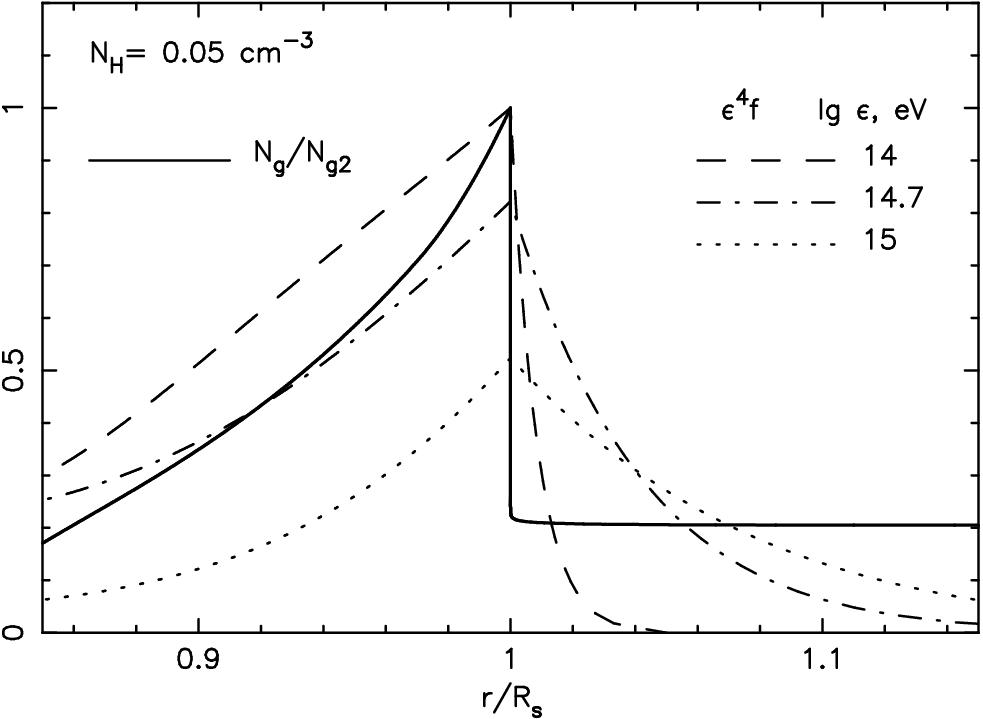}
 \caption{Radial profiles of the gas number density $N_\mathrm{g}$ (in units of
   the postshock number density $N_\mathrm{g2}$) and of the proton distribution
   $f$ in momentum $p$, multiplied by $\epsilon^4$, for three values of the
   proton energy $\epsilon$, in eV. The figure represents the actual dependence
   of the distribution function on energy, whereas the common normalization is
   arbitrary.}
\label{fig6}
\end{figure}

It should be noted that the cutoff of the theoretical \gr
spectrum (Fig.~\ref{fig5}) begins already at
$\epsilon_{\gamma}\approx 10^{13}$~eV. This value is unexpectedly low since
the overall proton power-law spectrum extends up to a proton energy $\epsilon
\approx 10^{15}$~eV (BKV09). The reason is the progressive decrease of overlap
between the radial gas density profile $N_\mathrm{g}(r)=\rho(r)/m_\mathrm{p}$
and the radial profile of the proton distribution function $f(r,\epsilon)$ at
energies $\epsilon > 10^{14}$~eV.  As Fig.~\ref{fig6} shows, the gas number
density has a peak $N_\mathrm{g2} =\sigma N_\mathrm{ISM}$ at the subshock
position $r=R_\mathrm{s}$ and decreases towards the remnant center by a factor
of 2 at a distance $l_\mathrm{g}\approx 0.06R_\mathrm{s}$ from the subshock
position. (The thickness of the thermal gas precursor is about
$10^{-4}R_\mathrm{s}$; therefore it is hardly visible in Fig.~\ref{fig6}.) The
corresponding radial dependence of the proton distribution in momentum $p$,
multiplied by $\epsilon^4$, i.e. of the spectral energy density $cp^4f(r,p)$,
with energies $\epsilon \approx cp \le 10^{14}$~eV, has a similar behaviour; it
shows a relatively small particle number in the upstream region
$r>R_\mathrm{s}$. At higher energies $\epsilon > 10^{14}$~eV the thickness of
the radial distribution increases with energy, so that for $\epsilon \approx
10^{15}$~eV its overlap with the gas density profile is by a factor of two
lower than for $\epsilon = 10^{14}$~eV. This leads to a lower cutoff of the \gr
spectrum compared with what one would expect in the simple case of
energy-independent overlap of CR protons with the gas distribution.
%

This spread of CRs into the upstream region $r>R_\mathrm{s}$, which for the
highest energies $\epsilon \sim \epsilon_\mathrm{max}(t)$ is faster than the
shock expansion $R_\mathrm{s}(t)$, represents the diffusive escape
of CRs from the expanding SNR and was predicted by
\citet{bk88}. Such a CR escape is expected in particular in the Sedov phase
when the maximum energy $\epsilon_\mathrm{max}(t)\propto B_0
R_\mathrm{s}V_\mathrm{s}$ of CRs, accelerated in the given evolutionary phase,
decreases with time, because the value of $\epsilon_\mathrm{max}$ is shock-size
limited \citep[see][for details]{ber96} rather than time limited. At any given
phase $t>t_0$ (where $t_0$ is the sweep-up time) efficient CR acceleration at
the SN shock takes place only for energies
$\epsilon\le\epsilon_\mathrm{max}(t)$, whereas for CRs with energies
$\epsilon_\mathrm{max}(t)<\epsilon <\epsilon_\mathrm{max}(t_0)$, produced
during earlier times, the acceleration process becomes inefficient and these
CRs expand into the upstream region. Within the SNR model discussed in this
paper CR escape is relatively slow because the model assumes Bohm diffusion in
the amplified magnetic field $B_0$ everywhere upstream for any CR energy. In
reality this is expected to be true only for CRs with energies $\epsilon
<\epsilon_\mathrm{max}(t)$ in the vicinity of the shock, where these CRs
produce significant magnetic field amplification. At large distances upstream
of the shock, $r-R_\mathrm{s}\gg 0.1R_\mathrm{s}$, or/and for higher CR
energies $\epsilon >\epsilon_\mathrm{max}(t)$, CR diffusion is much faster than
Bohm diffusion. Therefore in actual SNRs CR escape is expected to be faster and
more intense than the present model predicts \citep[see][for more detailed
considerations]{pz03,drury11}. Since SN~1006 is only at the very beginning of
the Sedov phase this underestimate of the magnitude of escape is however
not very critical.

\section{Summary}
Based on a nonlinear kinetic model for CR acceleration in SNRs the physical
properties of the remnant SN~1006 were examined in a detail
which corresponds to that of the available observations. Since the relevant
astronomical parameters as well as the synchrotron spectrum of SN~1006 are
measured with high accuracy, the values of the
relevant physical parameters of the model can be estimated with similar
accuracy for this SNR: proton injection rate $\eta\approx 2\times 10^{-4}$,
electron-to-proton ratio $K_{ep}\approx 3.8\times 10^{-4}$ and downstream
magnetic field strength $B_\mathrm{d}\approx 150$~$\mu$G.

As a result the flux of TeV emission detected by H.E.S.S. is predicted to
agree with an ISM hydrogen number density $
N_\mathrm{H}\approx 0.06~\mbox{cm}^{-3}$, consistent with the expectation from
X-ray measurements. The corresponding hydrodynamic SN explosion energy
$E_\mathrm{sn}\approx 1.9\times 10^{51}$~erg is somewhat above but fairly close
to the upper end $E_\mathrm{sn}= 1.6 \times 10^{51}$~erg of the typical range
of type Ia SN explosion energies. The efficiency of CR production is predicted
to be between 5 and 6 percent up to the present epoch and expected to approach
of the order of 10 precent during the further evolution. Normalized to a
  standard value $E_\mathrm{sn} = 10^{51}$~erg, already the present efficiency
  is about 10 percent.

The magnetic field amplification properties of this SNR
can be understood as the result of azimuthal variations of the nuclear ion
injection rate over the
projected SNR circumference in terms of injection at quasi-parallel shocks only
and the corresponding acceleration. This predicts the dipolar \gr emission
morphology and is compatible with a polar cap-type X-ray synchrotron
morphology. The \gr morphology is the key argument for the existence of an
energetically dominant nuclear CR component in SN 1006. The magnetic field
amplification allows the proton spectrum to extend to energies $\approx
10^{15}$~eV. The difference between the \gr emission from the NE polar region
and that from the SW polar region can be attributed to a small density
difference between those regions.

  It is shown that the high-energy tail of the \gr spectrum is affected by CR
  escape from the SNR interior, consistent with the comparatively low \gr
  cutoff near $10^{14}$~eV.

  It is also concluded that the radial profile of the TeV \gr emission measured
  by H.E.S.S. is evidence that the nuclear CR component is indeed efficiently
  produced. In the opposite case of inefficient nuclear CR production the
  magnetic field would not be expected to be amplified and the radial profile
  of the IC-dominated \gr emission would be expected to be significantly
  smoother than observed.

  The sum of all these results suggests the conclusion that SN~1006 is a
    CR source with a high efficiency of nuclear CR production,
  required for the Galactic CR sources, both in flux as well as in cutoff
  energy.


\acknowledgments
This work has been supported in part by the Russian Foundation for Basic
Research (grants 10-02-00154 and 11-02-12193) and by the Council of the
President of the Russian Federation for Support of Young Scientists and Leading
Scientific Schools (project No. NSh-1741.2012.2). EGB acknowledges the
hospitality of the Max-Planck-Institut f\"ur Kernphysik, where part of this
work was carried out.

\end{document}